\let\latexaddtocontents\addtocontents
\documentclass[a4paper,twocolumn,11pt]{quantumarticle}
\let\addtocontents\latexaddtocontents

\pdfoutput=1
\usepackage[utf8]{inputenc}
\usepackage[english]{babel}
\usepackage[T1]{fontenc}
\usepackage{amsmath}
\usepackage{hyperref}
\usepackage{tikz}
\usepackage{lipsum}
\usepackage{amsfonts, amssymb, amsmath}
\usepackage{physics}
\usepackage{stmaryrd}
\usepackage{listings}
\usepackage{appendix}
\usepackage[makeroom]{cancel}
\usepackage{mathtools}
\usepackage{float}
\usepackage{placeins}
\usepackage{hyperref}
\usepackage{array}
\usepackage{xcolor,colortbl}
\usepackage{booktabs}
\usepackage[backend=bibtex]{biblatex}
\newcommand\aug{\fboxsep=-\fboxrule\!\!\!\fbox{\strut}\!\!\!}

\newtheorem{defn}{Definition}[section]

\begin{document}

\title{Deep learning as a tool for quantum error reduction in quantum image processing}

\author{Krzysztof Werner}
\affiliation{Dept. of Computer Graphics, Vision and Digital Systems, Silesian University of Technology, Gliwice, Poland;}
\email{kwerner@polsl.pl}
\orcid{0000-0002-0327-837X}
\author{Kamil Wereszczyński}
\email{kamil.wereszczynski@polsl.pl}
\orcid{0000-0003-1686-472X}
\affiliation{Dept. of Computer Graphics, Vision and Digital Systems, Silesian University of Technology, Gliwice, Poland;}
\author{Rafał Potempa}
\orcid{0000-0002-0813-0606}
\author{Krzysztof Cyran}
\affiliation{Dept. of Computer Graphics, Vision and Digital Systems, Silesian University of Technology, Gliwice, Poland;}
\orcid{0000-0003-1789-4939}
\maketitle

\begin{abstract}
    Despite the limited availability and quantum volume of quantum computers, quantum image representation is a widely researched area.
    Currently developed methods use quantum entanglement to encode information about pixel positions. These methods range from using the angle parameter of the rotation gate (e.g., the Flexible Representation of Quantum Images, FRQI), sequences of qubits (e.g., Novel Enhanced Quantum Representation, NEQR), or the angle parameter of the phase shift gates (e.g., Local Phase Image Quantum Encoding, LPIQE) for storing color information.
    All these methods are significantly affected by decoherence and other forms of quantum noise, which is an inseparable part of quantum computing in the noisy intermediate-scale quantum era.
    These phenomena can highly influence the measurements and result in extracted images that are visually dissimilar to the originals. Because this process is at its foundation quantum, the computational reversal of this process is possible.
    There are many methods for error correction, mitigation, and reduction, but all of them use quantum computer time or additional qubits to achieve the desired result.
    We report the successful use of a generative adversarial network trained for image-to-image translation, in conjunction with Phase Distortion Unraveling error reduction method, for reducing overall error in images encoded using LPIQE.
\end{abstract}

\section{Introduction}

\par We propose the use of \textit{Generative Adversarial Networks} (GANs)  for error reduction, as a new method of error reduction in image processing, where a quantum computer is used to handle all calculations, and a classical computer is used for error reduction. In the future, we plan to implement the error reduction method on the quantum computer as well. 

\par This method is intended for use in quantum object detection of aviation instruments, in images taken on a Microsoft HoloLens 2 device inside the flight simulator cockpit.

\par \textbf{GAN} is a class of deep learning algorithm proposed in 2014 by Goodfellow et al. \cite{goodfellow2014generative}.
In this approach, two neural networks compete against each other in a zero-sum game as follows.
\begin{itemize}
% EDITOR COMMENT: OR "inputs"
    \item \label{g_model} \textbf{Generator}, produces candidates for output, using inputs original data.
    \item \label{d_model} \textbf{Discriminator}, evaluates candidates provided by the generator, recognizing whether the generated output is artificial or original. So, it is trained to distinguish generated output from the original samples. 
\end{itemize}
Such a game forces the generator to produce an output similar to the desired one. After the network is trained, it can be used to generate images or to evaluate image correctness.  Such networks have been used, for example, for generating paintings in the desired style from images.
The GAN networks can be implemented using a wide variety of programming libraries) e.g., TensorFlow \cite{tensorflow2015-whitepaper} and Keras \cite{chollet2015keras}, pygan or torchgan \cite{pal2021torchgan}).
% EDITOR COMMENT: OR "to generate images"
In our experiment, after training, we used the model for the generation of images with reduced error, based on the images reconstructed from the quantum computers.

% EDITOR COMMENT: Please choose which text should be formatted in bold.
\par (\textbf{LPIQE}) or the Local Phase Image Quantum Encoding method was proposed in 2020 \cite{wereszczynski_cosine_2020}.
It was meant to be implemented in photonic quantum computers (which are computers that use photons as the source of qubits).
Computers like this are currently in operation and are being offered to researchers.
The most known photonic processor from Xanadu, described by Madsen et al. in \cite{madsen_quantum_2022}, uses the squeezed states and measurement-based quantum computation model.
QuiX Quantum from the Netherlands (e.g., Somhorst et al. \cite{somhorst_quantum_2022}, Taballione et al. \cite{taballione_20-mode_2022} or de Goede et al. \cite{de_goede_high_2022}) uses other sources such as single photons or quantum dots as the supply of the quantum systems. 

LPIQE uses controlled phase shift gates for color representation. 
Such configurations are very natural for optic based systems, in contrary to other methods, that need to be implemented using significantly more optical devices.

For image representation, this method uses ${\lceil log_2 X \rceil+ \lceil log_2 Y \rceil+ 1}$ qubits, where \emph{X} and \emph{Y} references the X and Y dimensions of the encoded image, and the additional qubit is used for storing color information (in 1 channel mode).
For storing images represented in RGB spectrum, an additional 2 qubits are needed for R, G, or B channel selection.
Encoding color information in local phase allows the method to be combined with phase distortion unraveling (PDU) error reduction.

\par \textbf{Quantum error correction and reduction} is a widely researched area.
Current implementations of quantum computers are heavily affected by the impact of decoherence.
They require the temperatures of $0.015K$ (which is the consequence of the lack of room temperature superconductors) and a perfect vacuum.
% EDITOR COMMENT: "Earth" is the name of our planet. Hence, it is a proper noun and should start with a capital letter.
These conditions are nearly impossible to achieve on Earth, which makes quantum systems volatile to the slightest changes of the surrounding environment.
Decoherence is a term that encompasses the interactions, between the quantum computer and other particles, that are not expected by their operators, and which destroy the quantum nature of the system \cite{zurek_decoherence_2003}.

% EDITOR COMMENT: Use of singular "use" means that both quantum redundancy and measurement stabilization use the multidimensional Hilbert space, etc.
\par Some types of quantum error correction that aim to protect the qubit or a quantum system against decoherence and quantum noise are quantum redundancy and measurement stabilization, which use the multidimensional Hilbert space on which the quantum states are mapped \cite{bib:perfect_quantum_error_correction_code}, and
quantum surface codes, which are implemented on 2D qubit latices and comprehensively tackle the error of the whole quantum system \cite{bib:coherent_error_surface_codes}.
These methods of course have their shortcomings, such as using additional qubits or severely restricting the architecture of quantum circuits.
Either way, they are always implemented on a quantum computer alongside the algorithm.

\par Quantum error reduction focuses on minimizing the error.
Methods that are part of this family are implemented on classical machines, quantum computers, or as a hybrid--classical-quantum systems.
Some of the methods are Richardson extrapolation error mitigation \cite{richardson-extrapolation}, which performs measurement with artificially increased noise levels and then uses Richardson extrapolation to approximate no noise in the system and the quasi-probability method \cite{quasi-probability,near-future}, which tries to probabilistically inverse the noise process that occurred in the system.

\par The \textbf{PDU} method of error reduction was proposed in 2022 \cite{pdu}.
It was intended to be used in conjunction with quantum circuits using phase shift gates for calculations.
Error reduction is then conducted fully on a classical computer.
The PDU method is based on PDU functions.
These are interpolated functions for a set of points $(x_d, \varepsilon_d)$, where $x_d$ is the value for which the function was measured, and $\varepsilon_d$ is the recorded error.
The method is then applied on the result acquired from the quantum computer.

\begin{sloppypar} 
Given the current state of quantum hardware, the goal of our research is to overcome the quantum volume and noisy intermediate-scale quantum (NISQ) era limitations and to allow researchers to reach meaningful conclusions, even with imperfect systems.
This work shows experimentally that generative artificial intelligence could be used in error reduction in the NISQ era.
% EDITOR COMMENT: Your meaning of "transmitted state" was not clear. Please check the rewording.
In the area of quantum computing, the usefulness of generative AI can be conserved even beyond NISQ, because the fidelity of a transmitted state/of the transmitted state(s) is expected to be on a much lower level than inside the quantum computers (e.g., Tann \cite{tann_quantum_2022}).
\end{sloppypar}

\section{Materials and Methods}

\subsection{LPIQE method}

\par
\begin{figure}[t]
  \centering
  \includegraphics[width=0.4\textwidth]{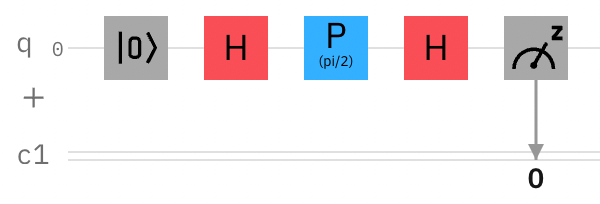}
  \caption{QCoSamp operator base of computation. Source: IBM's quantum composer.}
    \label{fig:q-co-samp}
\end{figure}

 LPIQE is the method of quantum image representation that uses the angle parameter of the controlled phase gate.
It was developed for implementation on quantum computers.
The basic building block of the method is the quantum cosine sampling (QCoSamp) operator, as shown in Figure \ref{fig:q-co-samp}.

\begin{figure}[t]
    \centering
    \includegraphics[width=0.5\textwidth]{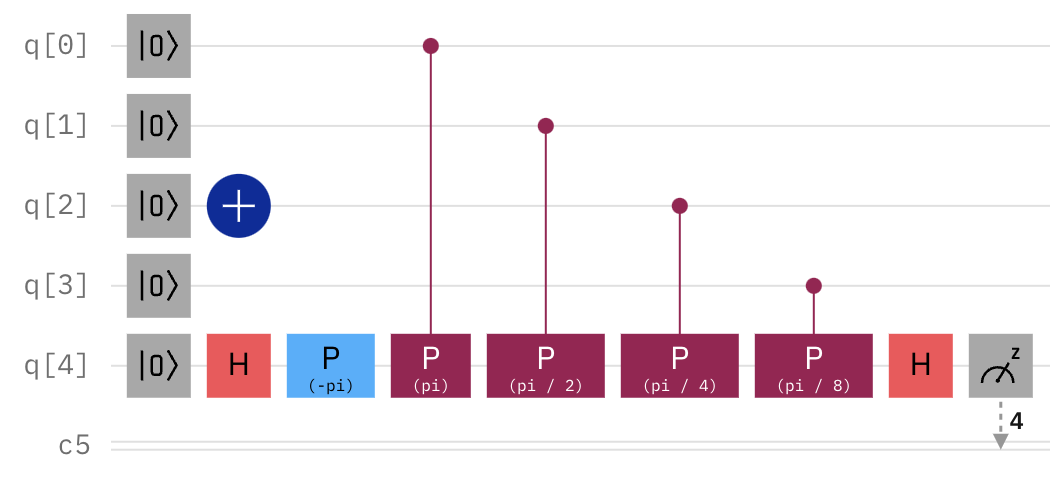}
    \caption{QCoSamp operator encoding the angle of $-\frac{3}{4}\pi$. Source: IBM's quantum composer.}
    \label{fig:q-co-samp-angle}
\end{figure}

The encoding circuit is composed of a base of computation and controlled phase shift gates.
The simplified version is shown in Figure \ref{fig:q-co-samp-angle}.

For the image encoding using this method, let's take the image of size H × W pixels.
It could be represented as a matrix $I_m=\big[\hat{p}_{r,c}\big]_{H\times W}$, where $\hat{p}_{r,c}$ is the intensity of the pixel placed in the $r$-th row and $c$-th column.
The image then can be flattened to a single vector using vertical vectorization to obtain 
\begin{equation}
    \overset{\longrightarrow}{I_m}=[p_j]^T, s.t.: j=rW+c.
\end{equation}

Then, it could be represented in the form of the state
\begin{gather}
\label{eq-image-state}
    \ket{I_m}=\big[e^{ip_0}\dots e^{ip_{J}},0^{M-J}\big]=\sum_{j=0}^Je^{ip_j}\ket{j}, \\J=WH-1, \;\; M=2^{\lceil log_2(WH)\rceil}\nonumber
\end{gather}
where the first part of the above state is the vector of $WH$ exponential functions of pixels intensities, and the second part is a complement to the quantum state's requirement of having the power of two coefficients. 

The final state can then be obtained by an operator defined by the matrix 
\begin{align}
    \Tilde{\mathcal{L}}(I_m)=\mathbf{1}\overset{\longrightarrow}{I_m}=&
    \begin{bmatrix}
            e^{ip_0}&0&\ldots&0\\
            0&e^{ip_1}&0\ldots&0\\
            \vdots&0&\ddots&0\\
            0&\ldots&0&e^{ip_J}\\
    \end{bmatrix}\nonumber\\
    \mathcal{L}(I_m)=&
    \begin{bmatrix}
            \Tilde{\mathcal{L}}(I_m)_{J\times J}&\aug&\mathbf{0}\\
            \hline
            \mathbf{0}&\aug&\mathbf{1}
    \end{bmatrix}_{M\times M}
\end{align}
where $\mathbf{1}$ is the identity matrix/operator.
At this point, the image can be easily encoded using a unitary gate from the matrix function present in most modern libraries.

\subsection{Error reduction by the PDU method}\label{error reduction}

\par PDU is the hybrid classical--quantum error reduction method.
It comprises of the set of PDU functions, that are calibrated for the quantum computer and circuit, and then interpolated between calibrated values.
The correlations between qubits are important for achieving the desired level of correction.

The calibration consists of running the function on a quantum computer and recording the differences between the expected values $\Tilde{\gamma}(x)$ and values received from the quantum computer $\gamma(x)$:

\begin{equation}
    \varepsilon(x)=\Tilde{\gamma}(x)-\gamma(x),
\end{equation}
where $x$ is an arbitrary value.
% EDITOR COMMENT: For correct grammar, a noun is required after "the set of those ___".
The function is interpolated for the set of those arguments.
These arguments are the probabilities following from local phases of qubits, extracted using the phase-kickback technique.
% EDITOR COMMENT: OR "in such a way that"
This can be generalized such that $x$ can be any object, which is described as follows:

\begin{defn}
Let $\ket{\psi}_x$ be the local phase $e^{ix}$ of the eigen-state $\ket{\psi}\in \mathbb{E}$ ($\mathbb{E}$ is a measurement basis) in a subspace of an $n$-qubit state, and $\mathfrak{p}\ket{\psi}_{x_k}$ is the experimentally designated probability amplitude for state $\ket{\psi}$ with phase $x_k$. In that case the function 
$$\varepsilon:\mathbb{E}\times [-\pi, \pi] \cap \mathbb{R}  \longrightarrow\mathbb{R}$$
is a \textbf{general PDU function} if and only if its each projection on the eigen-state is smooth due to phase and fulfills the following:
\begin{gather}
    \forall k:\;\; \varepsilon(\ket{\psi}, x_k)=\mathfrak{p}\ket{\psi}_{x_k}-\langle\psi\rangle_{x_k}\nonumber\\
    \forall\ket{\psi} \;\neg\exists \Tilde{\varepsilon}(\ket{\psi}, x):\nonumber\\
    \;\;\int_{-\pi}^\pi dx\;\; \Tilde{\varepsilon}(\ket{\psi}, x)
    <\int_{-\pi}^\pi dx\;\; \varepsilon(\ket{\psi}, x)
\end{gather}
\end{defn}

\subsection{GAN pix2pix network}\label{gan_networks}

\par GANs \cite{goodfellow2014generative} are a class of AI algorithms.
In this architecture, two networks compete with each other in a zero-sum game of two players -- the generator $G$ and the discriminator $D$. Given a random noise vector $z$ and ground truth $y$, the generator learns to produce artificial outputs similar to the ground truth:
\begin{equation}
\label{eq:gan_definition}
    G: z \to y.
\end{equation}
The \emph{PatchGAN} (or \emph{pix2pix}) model designed by Isola et al. \cite{isola_image_2017}, used for  the experiments in this work, belongs to a group of \emph{conditional GANs} proposed by Mirza et al. \cite{mirza2014conditional}. This means that except for the random noise $z$, the GAN takes a side input $x$ (e.g., the outline of an object used to generate its outputs):
\begin{equation}
\label{eq:pix2pix_definition}
    G: \{x, z\} \to y.
\end{equation}
Generally, the GAN's zero-sum game has an objective, also known as the value function:
\begin{equation}
\label{eq:gan_game}
    G^{*} = \arg \underset{G}{\min}\ \underset{D}{\max} V(G, D),
\end{equation}
\begin{multline}
\label{eq:value_function_gan}
    V(D, G) = 
    \mathbb{E}_{x\sim p_{\text{data}}}\log[D(x)] + \\
              \mathbb{E}_{x\sim p_{\text{model}}}\log[1 - D(G(z))],
\end{multline}
where $V$ is the value function, $x$ denotes the data, and $z$ is the input noise for the generator.
$D(x)$ represents the probability that $x$ came from the ground truth data, rather than from the model.
For a conditional GAN (e.g., pix2pix model), the value function \ref{eq:value_function_gan} takes the form of
\begin{multline}
\label{eq:value_function_conditional_gan}
    V(D, G) = 
    \mathbb{E}_{x\sim p_{\text{data}}}\log[D(x,y)] + \\
              \mathbb{E}_{x\sim p_{\text{model}}}\log[1 - D(x, G(x, z))].
\end{multline}
From  equations \ref{eq:value_function_gan} and \ref{eq:value_function_conditional_gan} comes the conclusion that the goal of the generator is to minimize the objective, while the goal of the discriminator is to maximize it. Namely, the two models compete with each other, where the discriminator tries to catch the generator's output, while the generator tries to outsmart it. The only difference is that the conditional GAN has a side input $x$, which is used by both the generator and discriminator. During training, the two models compete but at the convergence the generator's outputs are indistinguishable from the ground truth by the discriminator (i.e., $D(x) \approx 0.5$). The discriminator can be discarded during the model's inference phase \cite{goodfellow2016deep}.

\subsection{Experimental protocol}\label{sec-mm-exp-pro}

In the experiments, we explored the use of pix2pix GAN neural networks for error correction in quantum image processing.\\
We devised the following experiments:

\begin{enumerate}
% EDITOR COMMENT: OR "Reducing the errors in"
    \item \label{e1} Reducing the error in $16\times16$ images encoded with a quantum computer simulator, using images corrected with a PDU function as a reference.
    \item \label{e2} Reducing the error in $8\times8$ images encoded on real quantum computers, using images corrected with a PDU function as a reference.
    \item \label{e3} Reducing the error in $16\times16$ images encoded with a quantum computer simulator and corrected using a PDU function, with the original images as a reference.
    \item \label{e4} Reducing the error in $8\times8$ images encoded on real quantum computers and corrected using a PDU function, with the original images as a reference.
\end{enumerate}
Experiments \ref{e1} and \ref{e3} were done using the same initial dataset of 631 encoded and reconstructed images, and experiments \ref{e2} and \ref{e4} were done using the initial dataset of 61 encoded and reconstructed images. 

\begin{widetext}
%   \begin{equation}
% |\mathrm{AME}(n=6,q=5)\rangle=\sum_{i,j,k=0}^4 |i,j,k,i+j+k,i+2j+3k,i+3j+4k\rangle
%   \end{equation}
% \end{widetext}

    \begin{figure}[t]
        \centering
% EDITOR COMMENT: OR "used for the experiments". The meaning of "[the] experimental part" is not clear.
        \includegraphics[width=1\textwidth]{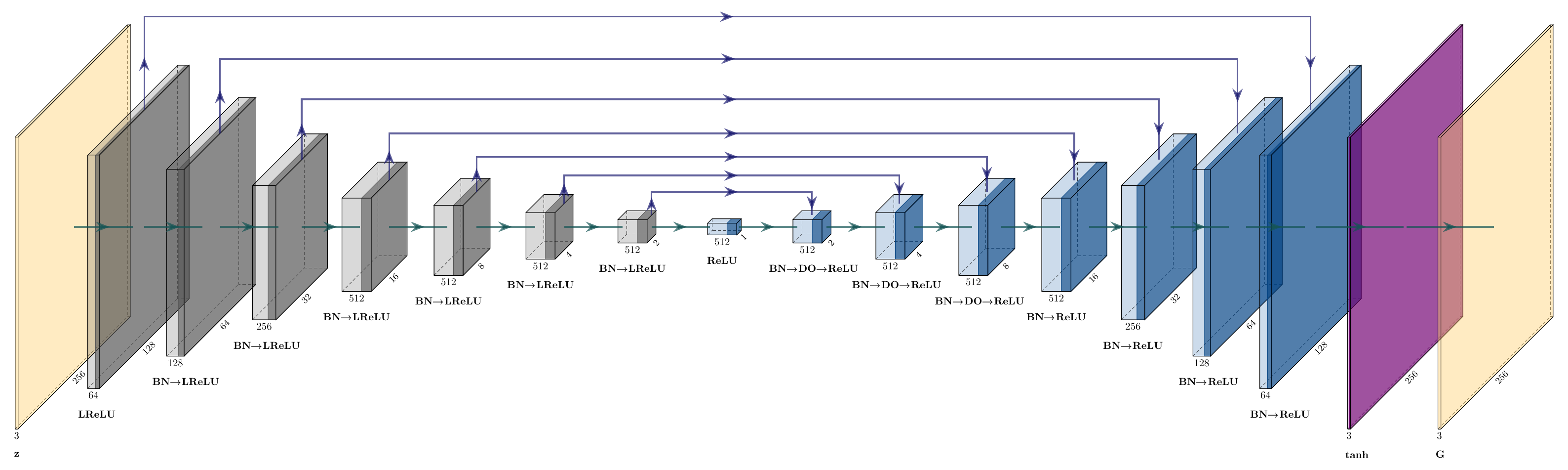}
        \caption{Overview of the generator network in the form of a U-Net used for the experimental part. All encoding layers' blocks consist of a convolutional layer (not shown), batch normalization (BN), and LeakyReLU (LReLU) activation. Decoding layers' blocks consist of a transposed convolutional layer (not shown), batch normalization (BN), dropout (DO), and ReLU activation. The output block is a transpose convolutional layer with a tanh function activation layer. The thickness dimension of each block corresponds to the number of its output channels, whereas its width corresponds to the edge of the resulting square image. Source: own compilation.}
        \label{fig:gan_generator2}
    \end{figure}

\end{widetext}

For the experiments, the GAN network was implemented using TensorFlow and Keras libraries for Python.
Both generator and discriminator were designed to accept $256\times256$ pixel images, to increase the networks' learning capacities.
The generator was implemented as a U-Net model with a total of 54,429,315 training parameters and 15 convolutional layers with LeakyReLU in the initial 7 encoding blocks. The bottleneck and decoding blocks used ReLU as the activation function. Between each of the consecutive encoding and decoding blocks, there was also a batch normalization operation. The first three decoding blocks additionally had a dropout with a value of 0.5. The network contained skip connections from encoders to corresponding decoders. An overview of the generator is shown in Figure \ref{fig:gan_generator2}.

\begin{figure}[t]
    \centering
% EDITOR COMMENT: OR "for the experiments"
    \includegraphics[width=0.4\textwidth]{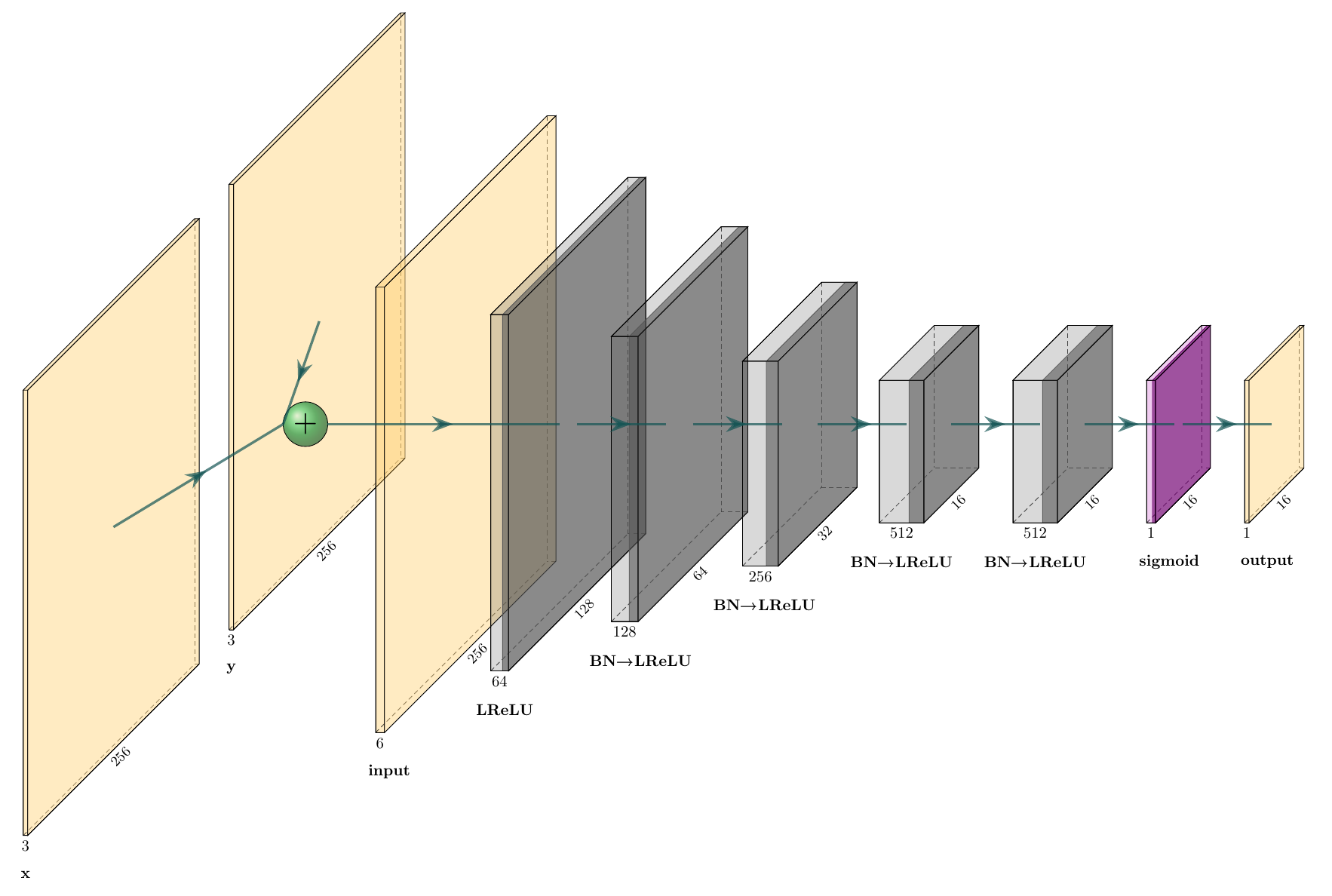}
    \caption{Overview of the discriminator network of pix2pix GAN used for the experimental part. All encoding layers' blocks consist of a convolutional layer (not shown), batch normalization (BN), and LeakyReLU (LReLU) activation. The output block is a convolutional layer with a sigmoid function activation layer. The thickness dimension of each block corresponds to the number of its output channels, whereas its width corresponds to the edge of the resulting square image. Source: own compilation.}
    \label{fig:gan_discriminator2}
\end{figure}

The discriminator was defined as a model with 6 convolutional layers, resulting in 6,968,257 training parameters. All the encoding blocks used LeakyReLU as the activation layers, and between all of them the batch normalization operation was performed. The output block used a sigmoid function as the activation layer. An overview of the generator is shown in Figure \ref{fig:gan_discriminator2}.


Each up- or down-scaling layer used a stride of $2\times2$. The convolutional and transpose convolutional layers used a kernel of size $4\times4$. For all LeakyReLU, the alpha parameter was set to 0.2.

The training was performed using an Adam optimizer \cite{kingma2014adam} with learning rate of $2\cdot10^{-4}$ and exponential decay rate for the first moment estimates $\beta_1 = 0.5$. The loss function was a weighted combination of binary cross-entropy defined as
\begin{equation}
\label{eq:binary_crossentropy}
    H(y, \hat{y}) = -y \log \hat{y} - (1 - y)\log(1 - \hat{y})
\end{equation}
and mean absolute error (MAE) defined as
\begin{equation}
\label{eq:mae}
    \text{MAE}(y, \hat{y}) = \frac{1}{N}|y - \hat{y}|,
\end{equation}
where $y$ denotes the true output and $\hat{y}$ its estimate, and N is the total number of performed estimations. The corresponding weights are equal to 1 and 100, respectively.

\subsubsection{Steps for the experiments}
\par Steps to reproduce the experiments are as follows:
\begin{enumerate}
    \item\label{step_1} Obtain images:
    \begin{enumerate}
        \item Implement the PDU function for the experiment, with the desired granularity of measurements (for the experiment, the granularity used was 5).
        \item Calibrate the PDU function.
        \item Generate black-and-white images (sizes for the experiment were $16\times16$ and $8\times$ pixels) with 256 shades of gray.
        \item Implement the quantum circuits for the generated images using LPIQE.
        \item Encode the images by running the circuits (for experiments \ref{e1} and \ref{e3}, the statevector\_simulator was used, and for \ref{e2} and \ref{e4}, IBMQ Nairobi was used).
        \item Reconstruct the images, and save the results alongside the original images.
        \item Use the calibrated PDU function on results to perform error reduction, and save the resulting image with the previous ones.
    \end{enumerate}
    \item\label{step_2} Create and train GAN network:
    \begin{enumerate}
        \item Create GAN neural network for picture-to-picture translation.
        \item Upscale the images to $256\times256$ pixels.
        \item Load training data appropriate for the experiment (for \ref{e1}, use the reconstructed images and original ones from the quantum simulator; for \ref{e2}, use the same but from the quantum computer; for \ref{e3}, use the images after error reduction and original ones from the quantum simulator; and for \ref{e4}, use the images after error reduction and original ones from a quantum computer).
        \item Divide the data into training and testing sets, and train the network on the training set (for experiments \ref{e1} and \ref{e3}, 250 epochs were performed, and for \ref{e2} and \ref{e4}, 1000 epochs were performed in training).
    \end{enumerate}
    \item\label{step_3} Evaluate the data:
    \begin{enumerate}
        \item For the testing set, collect the results from the generator.
        \item Descale the resulting images into original resolution.
        \item Compare the resulting pictures with the original ones.
    \end{enumerate}
\end{enumerate}

% ###############################

\section{Results}
\subsection{Reducing the error in $16\times16$ images encoded using a quantum simulator, using original images as a reference.}\label{sim_original} 
For the experiment, we collected 631 distinct images reconstructed after encoding on a quantum simulator.

\begin{figure}[t]
    \centering
    \begin{tabular}{ccc}
         \includegraphics[width=0.1\textwidth]{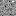}&
         \includegraphics[width=0.1\textwidth]{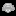}&
         \includegraphics[width=0.1\textwidth]{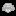}
    \end{tabular}
% EDITOR COMMENT: OR "represents the reconstructed image after encoding on a simulator (left), "
    \caption{Example results for the image encoded using the statevector simulator. The figure represents the image after encoding on a simulator and reconstruction (left), the image generated by the neural network (middle), and the expected original image (right).}
    \label{fig:e1-example}
\end{figure}

An example of the correction performed is shown in Figure \ref{fig:e1-example}. 
For that particular example, the Pearson's correlation increased from -0.01935 to 0.99962 after using the generator from the GAN network.

\begin{figure}[t]
    \centering
    \begin{tabular}{c}
        \includegraphics[width=0.4\textwidth]{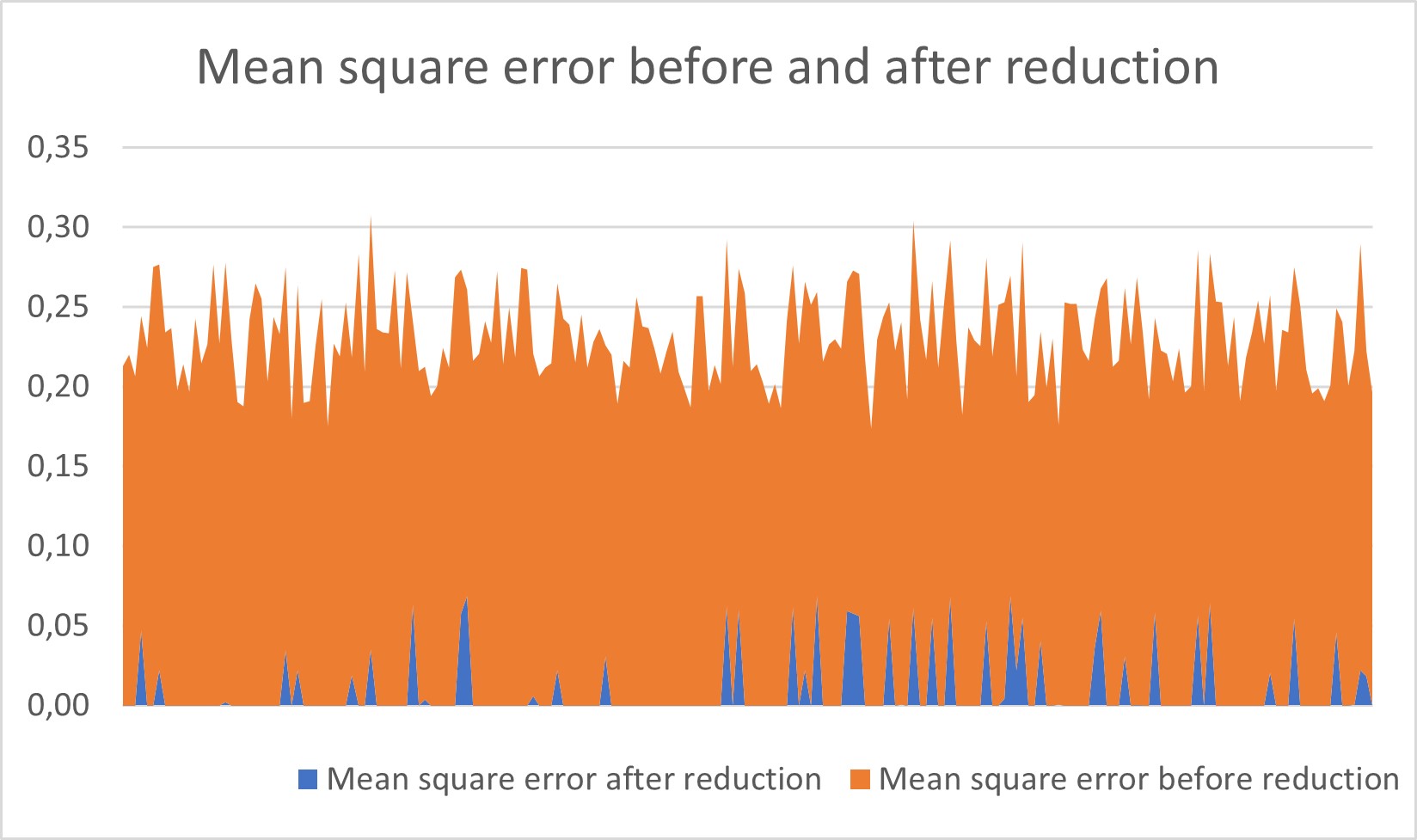}
        
    \end{tabular}
    \caption{The chart shows the difference in mean square error before and after reduction.}
    \label{fig:e1-mse}
\end{figure}

\begin{figure}[t]
    \centering
    \begin{tabular}{c}
        \includegraphics[width=0.4\textwidth]{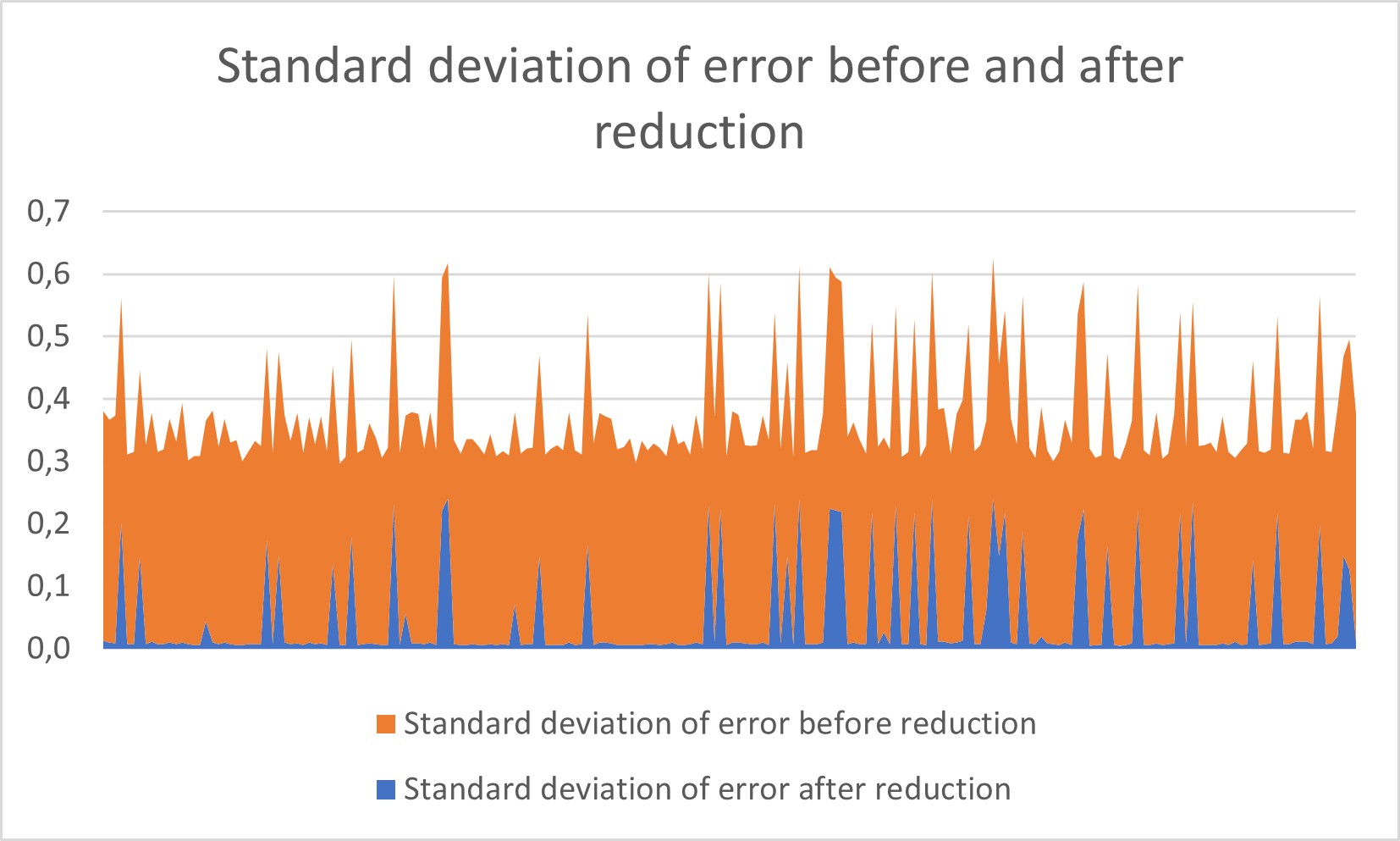}
        
    \end{tabular}
    \caption{The chart shows the difference in standard deviation of error before and after reduction.}
    \label{fig:e1-std-dev}
\end{figure}

Detailed statistics are shown in Figures \ref{fig:e1-mse}, \ref{fig:e1-std-dev} and Tables \ref{tab:experiment_1_correlations}, \ref{tab:experiment_1_correlations_2}.

\begin{table}[t]
    \caption{Minimal, maximal, and mean correlation coefficients values, MSE, and STDEV of error for experiment \ref{e1} before error reduction.}
    \centering
        \begin{tabular}{c|c|c|c}
        \toprule
        & min & mean & max \\
        \midrule
        Pearson's R & -0.15957 & 0.00708 & 0.16703 \\
        Mean square error & 0.17357 & 0.22277 & 0.28308 \\
        Std dev of error & 0.29144 & 0.32861 & 0.38558
        \end{tabular}
    \label{tab:experiment_1_correlations}
\end{table}

\begin{table}[t]
    \caption{Minimal, maximal, and mean correlation coefficients values, MSE, and STDEV of error for experiment \ref{e1} after error reduction.}
    \centering
        \begin{tabular}{c|c|c|c}
        \toprule
        & min & mean & max  \\
        \midrule
        Pearson's R & 0.72075 & 0.96135 & 0.99993 \\
        Mean square error & 0.00003 & 0.00885 & 0.06843 \\
        Std dev of error & 0.00484 & 0.04444 & 0.24108
        \end{tabular}
    \label{tab:experiment_1_correlations_2}
\end{table}

For this experiment, we observed a numerical improvement in all measured statistics.
The standard deviation of error dropped on average 7.4 times, and the mean square error dropped on average 25.1 times.
Pearson's R also rose by 0.954.

\subsection{Reducing the error in $8\times8$ images encoded on real quantum computers, using images corrected with a PDU function as a reference.} 
For the experiment, we collected 61 distinct images reconstructed after encoding on a quantum computer and fed them to the neural network to obtain images with reduced error.

\begin{figure}[t]
    \centering
    \begin{tabular}{ccc}
         \includegraphics[width=0.1\textwidth]{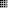}&
         \includegraphics[width=0.1\textwidth]{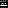}&
         \includegraphics[width=0.1\textwidth]{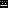}
    \end{tabular}
    \caption{Example results for the image encoded using an IBMQ Nairobi, 7 qubit quantum computer. The figure represents the image after encoding and reconstruction (left), the image generated by the neural network (middle), and the expected original image (right).}
    \label{fig:e2-example}
\end{figure}

An example is shown in Figure \ref{fig:e2-example}. 
For that particular example, the Pearson's correlation increased from 0.24326 to 0.89631.

\begin{figure}[t]
    \centering
    \begin{tabular}{c}
        \includegraphics[width=0.4\textwidth]{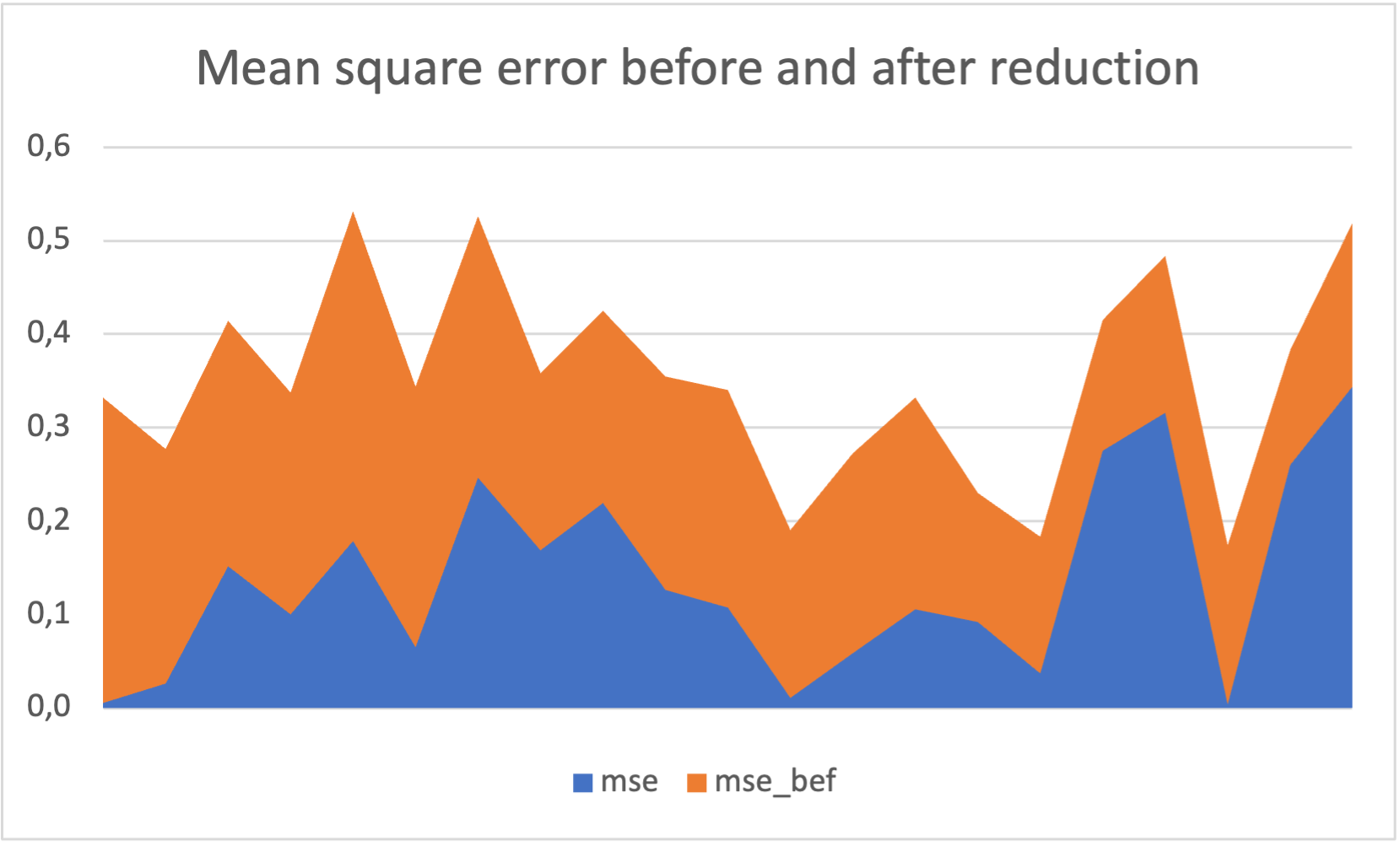}
    \end{tabular}
% EDITOR COMMENT: OR "for experiments performed". You may also state the experiment number explicitly (e.g., "for experiment 2 performed ..."
    \caption{The chart shows the difference in mean square error for the experiment performed on a real quantum device.}
    \label{fig:e2-mse}
\end{figure}

\begin{figure}[t]
    \centering
    \begin{tabular}{c}
        \includegraphics[width=0.4\textwidth]{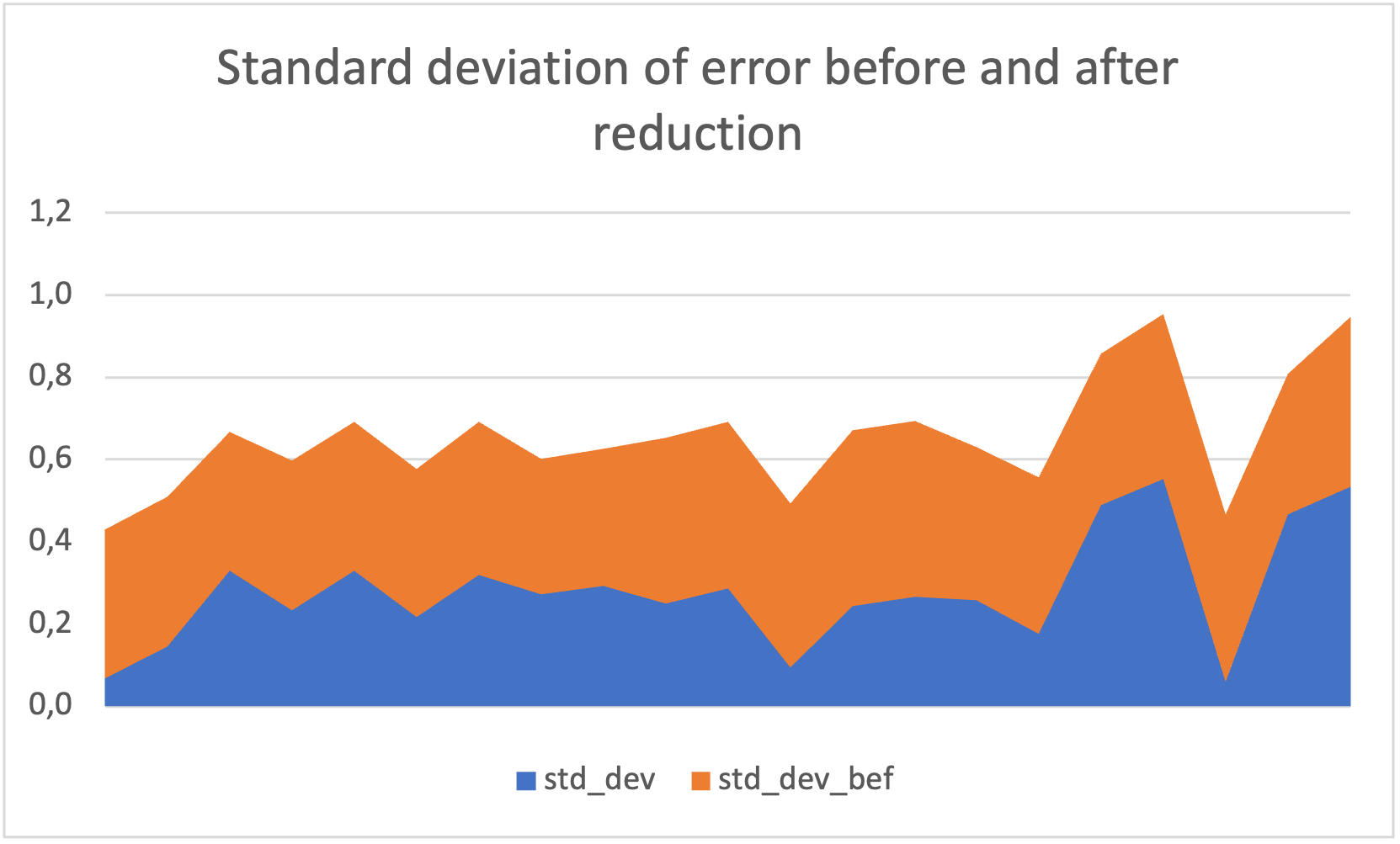}
    \end{tabular}
% EDITOR COMMENT: OR "for experiments performed"
    \caption{The chart shows the difference in standard deviation of error for the experiment performed on a real quantum device.}
    \label{fig:e2-std-dev}
\end{figure}

The statistics for all collected samples are shown in Tables \ref{tab:experiment_2_correlations}, \ref{tab:experiment_2_correlations_2}, and Figures \ref{fig:e2-mse}, \ref{fig:e2-std-dev}.

\begin{table}[t]
    \caption{Minimal, maximal, and mean correlation coefficients values, MSE, and STDEV of error for experiment \ref{e2} before error reduction.}
    \centering
        \begin{tabular}{c|c|c|c}
        \toprule
        & min & mean & max \\
        \midrule
        Pearson's R & 0.01121 & 0.17262 & 0.31465 \\
        Mean square error & 0.12317 & 0.21375 & 0.35261 \\
        Std dev of error & 0.32898 & 0.37183 & 0.42898
        \end{tabular}
    \label{tab:experiment_2_correlations}
\end{table}

\begin{table}[t]
    \caption{Minimal, maximal, and mean correlation coefficients values, MSE, and STDEV of error for experiment \ref{e2} after error reduction.}
    \centering
        \begin{tabular}{c|c|c|c}
        \toprule
        & min & mean & max  \\
        \midrule
        Pearson's R & -0.32539 & 0.56511 & 0.98399 \\
        Mean square error & 0.00410 & 0.10783 & 0.34393 \\
        Std dev of error & 0.05883 & 0.26588 & 0.55277
        \end{tabular}
    \label{tab:experiment_2_correlations_2}
\end{table}

For this experiment, we still observed a numerical improvement in all measured statistics, although the level of correction was much lower.
The standard deviation of error dropped on average 1.45 times, whereas the mean square error dropped on average 1.26 times. Pearson's R also rose by 0.392 to 0.565 with p-value of 0.005.

\subsection{Reducing the error in $16\times16$ images encoded with a quantum computer simulator and corrected using a PDU function, with the original images as a reference.} 
\begin{figure}[t]
    \centering
    \begin{tabular}{ccc}
         \includegraphics[width=0.1\textwidth]{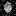}&
         \includegraphics[width=0.1\textwidth]{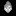}&
         \includegraphics[width=0.1\textwidth]{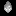}
    \end{tabular}
    \caption{Example results for the image encoded using the statevector simulator. The figure represents the image after encoding on a simulator, reconstruction, and error reduction with a PDU method (left), the image generated by the neural network (middle), and the expected original image (right).}
    \label{fig:e3-example}
\end{figure}

In this experiment, we used images encoded onto a quantum simulator, decoded and then corrected with a PDU function, and fed to the GAN network.
An example is shown in Figure \ref{fig:e3-example}.

\begin{figure}[t]
    \centering
    \begin{tabular}{c}
        \includegraphics[width=0.4\textwidth]{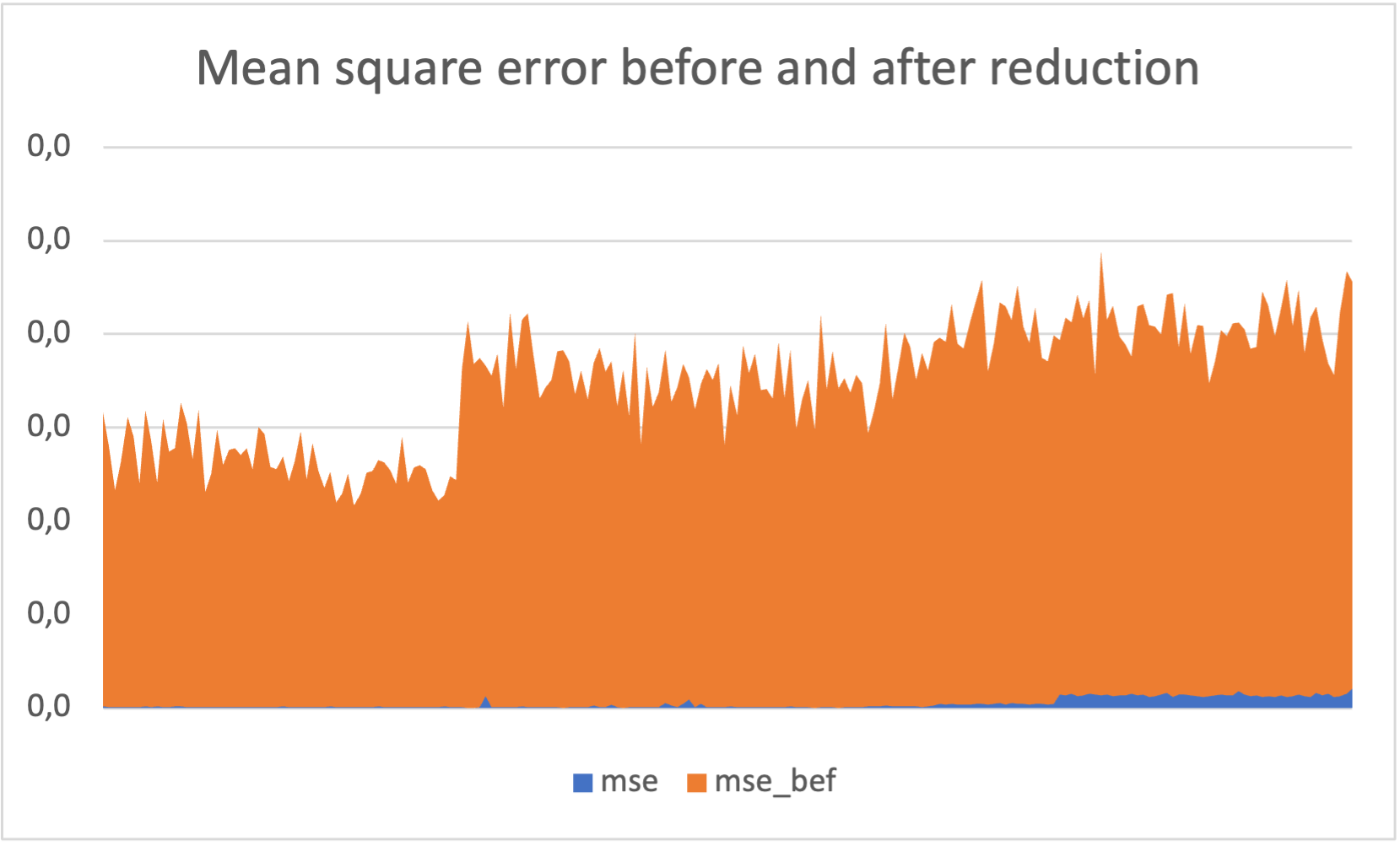}
    \end{tabular}
    \caption{The chart shows the difference in mean square error before and after error reduction.}
    \label{fig:e3-mse}
\end{figure}

\begin{figure}[t]
    \centering
    \begin{tabular}{c}
        \includegraphics[width=0.4\textwidth]{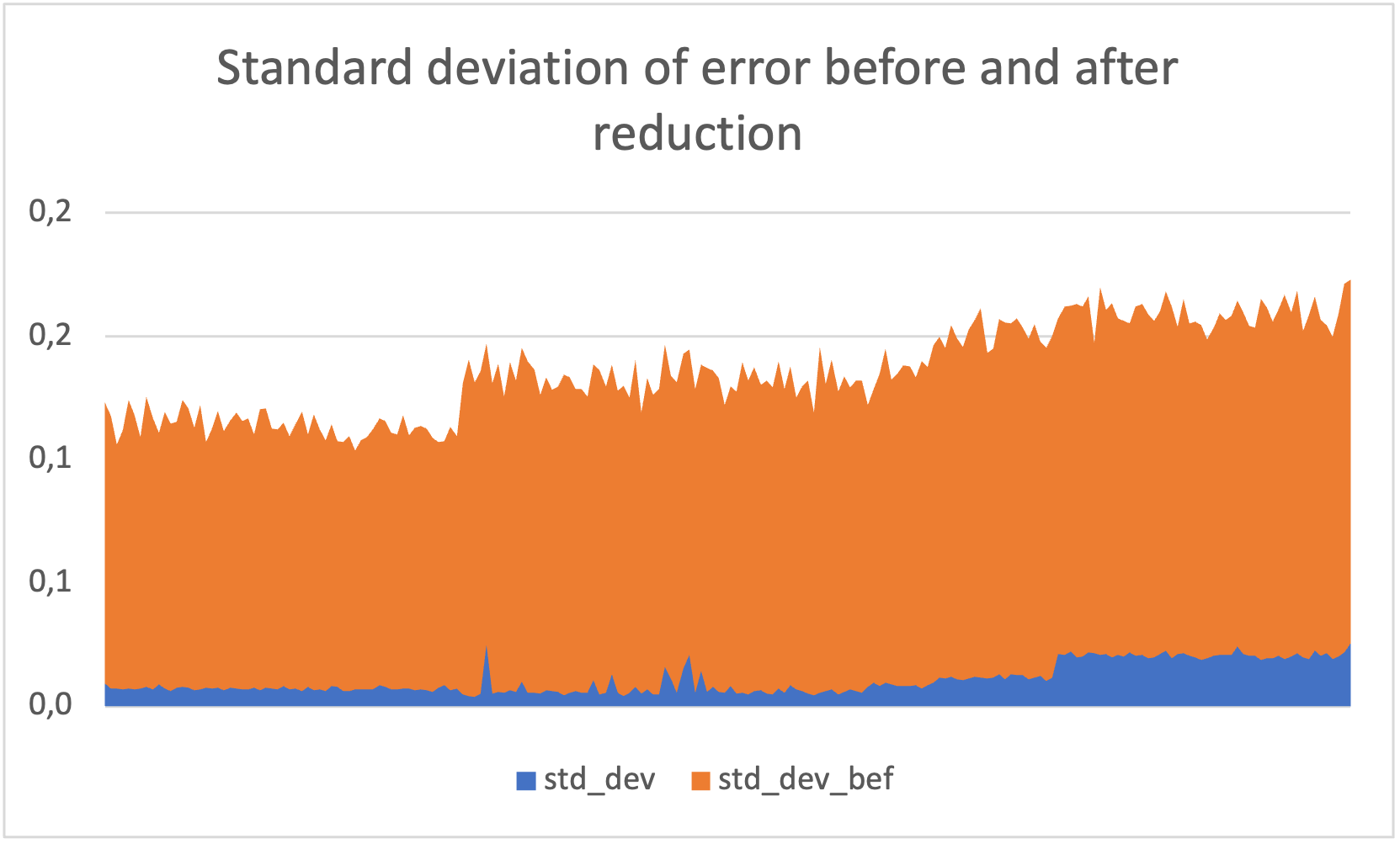}
    \end{tabular}
    \caption{The chart shows the difference in standard deviation of error before and after error reduction.}
    \label{fig:e3-std-dev}
\end{figure}

Detailed statistics are shown in Figures \ref{fig:e3-mse}, \ref{fig:e3-std-dev} and Tables \ref{tab:experiment_3_correlations}, \ref{tab:experiment_3_correlations_2}.

\begin{table}[t]
    \caption{Minimal, maximal, and mean correlation coefficients values, MSE, and STDEV of error for experiment \ref{e3} before error reduction.}
    \centering
        \begin{tabular}{c|c|c|c}
        \toprule
        & min & mean & max \\
        \midrule
        Pearson's R & 0.87700 & 0.91805 & 0.94214 \\
        Mean square error & 0.01077 & 0.01779 & 0.02378 \\
        Std dev of error & 0.09665 & 0.12655 & 0.14971
        \end{tabular}
    \label{tab:experiment_3_correlations}
\end{table}

\begin{table}[t]
    \caption{Minimal, maximal, and mean correlation coefficients values, MSE, and STDEV of error for experiment \ref{e3} after error reduction.}
    \centering
        \begin{tabular}{c|c|c|c}
        \toprule
        & min & mean & max  \\
        \midrule
        Pearson's R & 0.99639 & 0.99983 & 0.99993 \\
        Mean square error & 0.000014 & 0.00007 & 0.00106 \\
        Std dev of error & 0.00371 & 0.00769 & 0.02574
        \end{tabular}
    \label{tab:experiment_3_correlations_2}
\end{table}

In this experiment, we observed a numerical improvement in Pearson's R test, which rose by 0.0818 to 0.999 with p-value of 3.25e-83.
Meanwhile, the mean square error dropped 270,14 times, whereas the standard deviation of error dropped 16,46 times.
\par For the collected mean square error data, we also calculated the standard deviation for mean square error ($\sigma$) and calculated statistics about the percentage of data that fit the equation $mse\pm n * \sigma$. We found that for $n = 1$, 75.71\% of data fell into this category; for $n = 2$, it was 97.62\%; and for $n = 3$, it was 99.52\%.
We also calculated that for 95\% of the data to meet the criteria of the equation, $n$ should be equal to 1.845; for 90\%, $n$ should be equal to 1.716; for 75\%, $n$ should be equal to 0.756; and for 50\%, $n$ should be equal to 0.679.

\subsection{Reducing the error in $8\times8$ images encoded on real quantum computers and corrected using a PDU function, with the original images as a reference.} 
\begin{figure}[t]
    \centering
    \begin{tabular}{ccc}
         \includegraphics[width=0.1\textwidth]{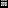}&
         \includegraphics[width=0.1\textwidth]{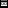}&
         \includegraphics[width=0.1\textwidth]{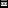}
    \end{tabular}
    \caption{Example results for the image encoded using an IBMQ Nairobi, 7 qubit quantum computer. The figure represents the image after encoding, decoding, and correction with a PDU method (left), the image generated by the neural network (middle), and the expected original image (right).}
    \label{fig:e4-example}
\end{figure}

In this experiment, to reduce the error after encoding and decoding the image on a quantum computer, first the PDU method was used, and then the trained GAN network was used to reduce the error further.
An example of such a correction is shown in Figure \ref{fig:e4-example}.

\begin{figure}[t]
    \centering
    \begin{tabular}{c}
        \includegraphics[width=0.4\textwidth]{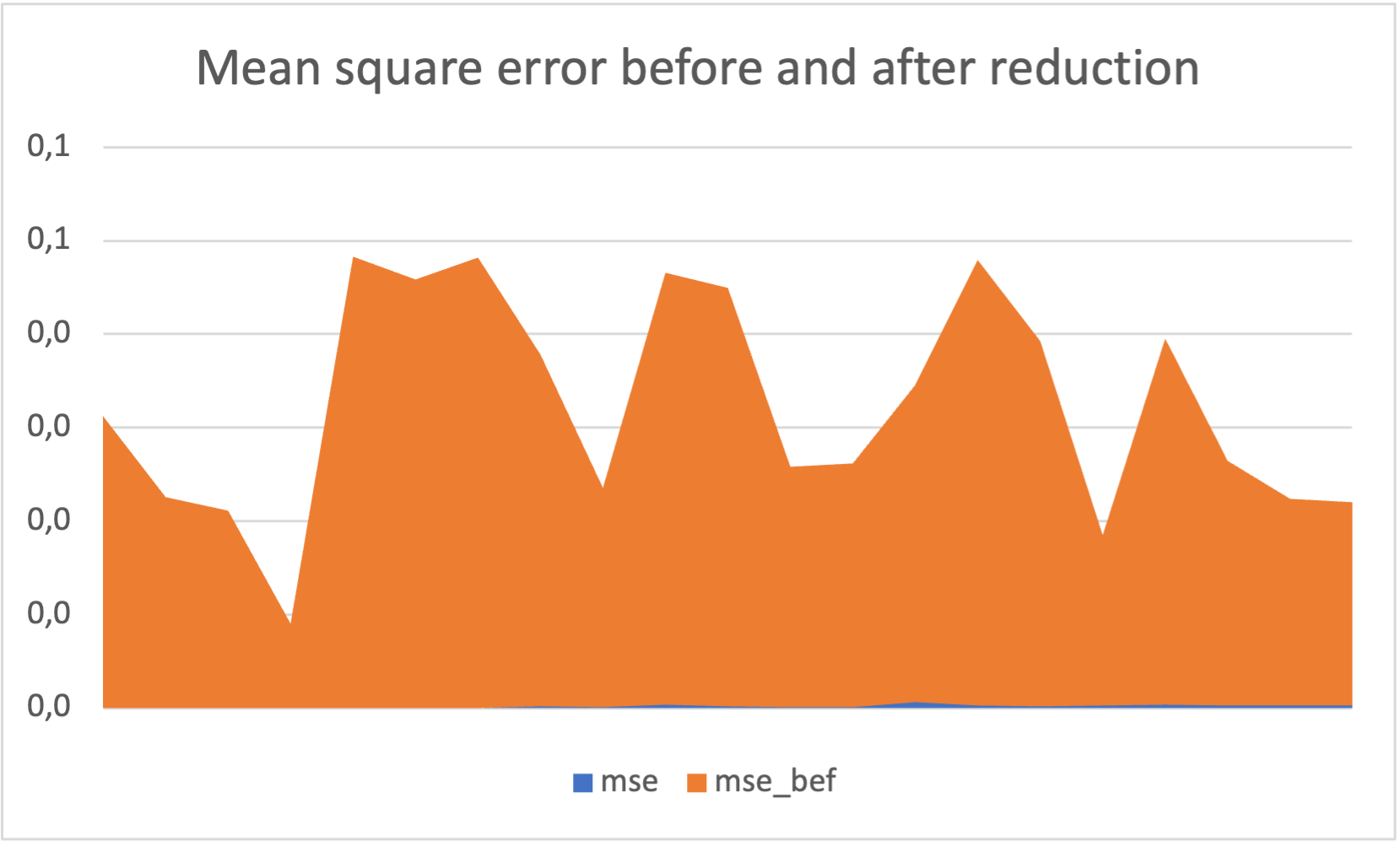}
    \end{tabular}
% EDITOR COMMENT: OR "for experiments performed"
    \caption{The chart shows the difference in mean square error before and after reduction for the experiment performed on a real quantum computer.}
    \label{fig:e4-mse}
\end{figure}

\begin{figure}[t]
    \centering
    \begin{tabular}{c}
        \includegraphics[width=0.4\textwidth]{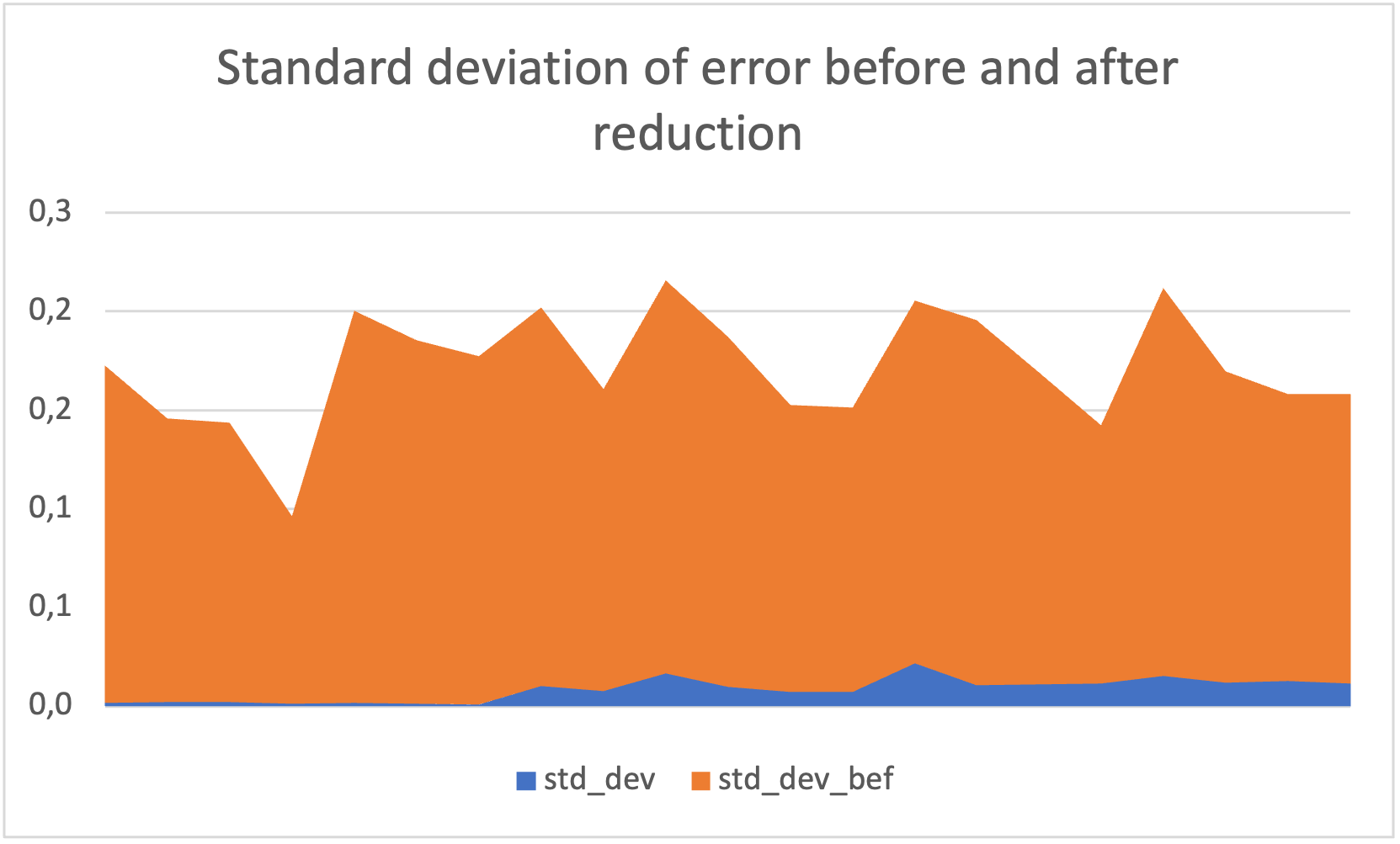}
    \end{tabular}
% EDITOR COMMENT: OR "for experiments performed"
    \caption{The chart shows the difference in standard deviation of error before and after reduction for the experiment performed on a real quantum computer.}
    \label{fig:e4-std-dev}
\end{figure}

The statistics for all collected samples are shown in Tables \ref{tab:experiment_4_correlations}, \ref{tab:experiment_4_correlations_2}, and Figures \ref{fig:e4-mse}, \ref{fig:e4-std-dev}.

\begin{table}[t]
    \caption{Minimal, maximal, and mean correlation coefficients values, MSE, and STDEV of error for experiment \ref{e4} before error reduction.}
    \centering
        \begin{tabular}{c|c|c|c}
        \toprule
        & min & mean & max \\
        \midrule
        Pearson's R & 0.79338 & 0.88266 & 0.96455 \\
        Mean square error & 0.00906 & 0.03122 & 0.04825 \\
        Std dev of error & 0.09512 & 0.15817 & 0.19945
        \end{tabular}
    \label{tab:experiment_4_correlations}
\end{table}

\begin{table}[t]
    \caption{Minimal, maximal, and mean correlation coefficients values, MSE, and STDEV of error for experiment \ref{e4} after error reduction.}
    \centering
        \begin{tabular}{c|c|c|c}
        \toprule
        & min & mean & max  \\
        \midrule
        Pearson's R & 0.99805 & 0.99970 & 1.00000 \\
        Mean square error & 0.0000006 & 0.00016 & 0.00065 \\
        Std dev of error & 0.00069 & 0.00993 & 0.02175
        \end{tabular}
    \label{tab:experiment_4_correlations_2}
\end{table}

In terms of measured statistics, we observed further improvement noticed mostly in the drop of mean square error, which was 226 times smaller after using the neural network for error reduction. Also, the standard deviation of error dropped by 15.9 times, and the average Pearson's R rose by 0.117 to the value of 0.9997 with p-value of 6.2e-78.
\par For the collected mean square error data, we also calculated the standard deviation for mean square error ($\sigma$) and calculated statistics about the percentage of data that fit the equation $mse\pm n * \sigma$. We found that for $n = 1$, 85.71\% of data fell into this category; for $n = 2$, it was 95.23\%; and for $n = 3$, it was 100\%.
We have calculated that for 95\% of the data to meet the criteria of the equation, $n$ should be equal to 1.168; for 90\%, $n$ should be equal to 1.001; for 75\%, $n$ should be equal to 0.975; and for 50\%, $n$ should be equal to 0.877.

We suspect that this noticeable error reduction was possible because the PDU function already corrected the bulk of errors induced by quantum noise, leaving the result less random (compared to the original) and making it easier for the neural network to observe patterns.
% EDITOR COMMENT: It is not clear which table you are referring to by "the observed statistics for the standard deviation of the combined mean square error". This is because the phrase "combined mean square error" does not appear anywhere else in the paper. It would also be correct grammar (but perhaps not correct meaning) to say, "seemed to be higher in experiment 3." Please reword.
\par Also, the observed statistics for the standard deviation of the combined mean square error seemed to be higher than those obtained in experiment \ref{e3}.
We suspect that the reason for that could be the difference in quantum computer architecture for which the simulator was constructed.
The simulator was made for simulating ion trap computers, and phase elements of qubits seemed to be simulated in a less precise way.

%%%%%%%%%%%%%%%%%%%%%%%%%%%%%%%%%%%%%%%%%%
\section{Discussion \& Conclusions}

In this paper, we showed that GANs can be used to further reduce the error from quantum noise in quantum calculations that use quantum sampling.
Existing error reduction methods (such as PDU presented in this paper) can lower the quantum error significantly enough for neural networks to be effectively used to further reduce the error to a point where visual identification of objects on the picture is possible.
With the further development of quantum computers, encoded images will be significantly larger, which would require the neural network to be more complex. This would require special equipment to be able to train them effectively.
Although we proved that GAN networks can reduce the quantum noise error of images, they are much less effective, especially on normal computers.
Perhaps with the use of supercomputers, this method would be able to reduce the quantum error in a similar fashion, but as researchers, we did not have access to such a machine.
The combined PDU and GAN error reduction was statistically significant, which was confirmed by visual similarity between the expected result and one produced by the neural network.
% EDITOR COMMENT: OR "The Pearson coefficient was on the level of $0.99$ with the p-value near zero, "
The Pearson coefficients were on the level of $0.99$ with p-values near zero, which proves the very high linear correlation between the original and corrected images.
We are especially happy with the results obtained from the real quantum computers in experiment \ref{e4}.
The mean square error of 0.00014 and standard deviation of error of 0.00993 showed that this method could be used with existing, imperfect quantum computers from the NISQ era and produce results that -- at least for us -- are visually indistinguishable from the original images.
The results obtained from experiments \ref{e1} and \ref{e3} also showed that these networks can be used beyond the NISQ era, for quantum error reduction.
We would like to develop this method further so it can be used as part of a quantum image processing pipeline, where images will be encoded onto a quantum computer, and then measured and decoded, and image transformations performed in the encoded state.
We focused on quantum image processing because we are interested in developing methods in this area. However, these methods could be used for other types of data.
In conclusion, this paper presented how neural networks can enhance the existing PDU error reduction method for quantum encoded images. 
The method will be utilized in the image processing in photonic quantum solutions and for quantum communication in faster-then-light communication.
It could also be used to pursue the goal of quantum object detection. This could be utilized in detecting aviation instruments on pictures taken inside a flight simulator cockpit, which could be used for pilot training.

% The quantumarticle document class comes bundled with a \href{https://raw.githubusercontent.com/quantum-journal/quantum-journal/master/quantum-lyx-template.lyx}{LyX layout} that allows to typeset manuscripts with the LyX document processor instead of directly writing LaTeX code. Please be aware that this is a beta feature that might not receive the same level of support as the quantumarticle document class itself.

\section{Version}
\label{sec:version}
This is quantumarticle version v\quantumarticleversion.

\printbibliography

@article{wereszczynski_cosine_2020,
	title = {Cosine series quantum sampling method with applications in signal and image processing},
	url = {http://arxiv.org/abs/2011.12738},
	urldate = {2021-06-20},
	journal = {arXiv:2011.12738 [quant-ph]},
	author = {Wereszczyński, Kamil and Michalczuk, Agnieszka and Pęszor, Damian and Paszkuta, Marcin and Cyran, Krzysztof and Polański, Andrzej},
	month = {11},
	year = {2020},
	note = {arXiv: 2011.12738},
}

@inproceedings{isola_image_2017,
    title={Image-to-image translation with conditional adversarial networks},
    author={Isola, Phillip and Zhu, Jun-Yan and Zhou, Tinghui and Efros, Alexei A},
    booktitle={Proceedings of the IEEE conference on computer vision and pattern recognition},
    pages={1125--1134},
    year={2017}
}

@book{goodfellow2016deep,
  title={Deep learning},
  author={Goodfellow, Ian and Bengio, Yoshua and Courville, Aaron},
  year={2016},
  publisher={MIT press}
}

@article{pdu,
    author = {Werner, Krzysztof and Wereszczy\'{n}ski, Kamil and Michalczuk, Agnieszka},
    title = {Experiment-Driven Quantum Error Reduction},
    year = {2022},
    isbn = {978-3-031-08759-2},
    publisher = {Springer-Verlag},
    address = {Berlin, Heidelberg},
    url = {https://doi.org/10.1007/978-3-031-08760-8{\_17}},
    doi = {10.1007/978-3-031-08760-8{\_17}},
    journal = {Computational Science – ICCS 2022: 22nd International Conference, London, UK, June 21–23, 2022, Proceedings, Part IV},
    pages = {195–201},
    numpages = {7},
}

@article{zurek_decoherence_2003,
	title = {Decoherence, einselection, and the quantum origins of the classical},
	volume = {75},
	issn = {0034-6861, 1539-0756},
	url = {http://arxiv.org/abs/quant-ph/0105127},
	doi = {10.1103/RevModPhys.75.715},
	number = {3},
	urldate = {2021-07-26},
	journal = {Reviews of Modern Physics},
	author = {Zurek, Wojciech H.},
	month = {5},
	year = {2003},
	note = {arXiv: quant-ph/0105127},
	pages = {715--775},
}

@article{taballione_20-mode_2022,
	title = {20-{Mode} {Universal} {Quantum} {Photonic} {Processor}},
	copyright = {arXiv.org perpetual, non-exclusive license},
	url = {https://arxiv.org/abs/2203.01801},
	doi = {10.48550/ARXIV.2203.01801},
	urldate = {2022-08-23},
        journal={Quantum},
	author = {Taballione, Caterina and Anguita, Malaquias Correa and de Goede, Michiel and Venderbosch, Pim and Kassenberg, Ben and Snijders, Henk and Kannan, Narasimhan and Smith, Devin and Epping, Jörn P. and van der Meer, Reinier and Pinkse, Pepijn W. H. and Vlekkert, Hans van den and Renema, Jelmer J.},
	year = {2022},
	note = {Publisher: arXiv
Version Number: 3},
}

@misc{de_goede_high_2022,
	title = {High {Fidelity} 12-{Mode} {Quantum} {Photonic} {Processor} {Operating} at {InGaAs} {Quantum} {Dot} {Wavelength}},
	url = {http://arxiv.org/abs/2204.05768},
	urldate = {2022-08-23},
	publisher = {arXiv},
	author = {de Goede, Michiel and Snijders, Henk and Venderbosch, Pim and Kassenberg, Ben and Kannan, Narasimhan and Smith, Devin H. and Taballione, Caterina and Epping, Jörn P. and Vlekkert, Hans van den and Renema, Jelmer J.},
	month = {4},
	year = {2022},
	note = {arXiv:2204.05768 [physics, physics:quant-ph]},
}

@article{somhorst_quantum_2022,
	title = {Quantum photo-thermodynamics on a programmable photonic quantum processor},
	copyright = {arXiv.org perpetual, non-exclusive license},
    journal={arXiv preprint arXiv:2201.00049 v1},
	url = {https://arxiv.org/abs/2201.00049},
	doi = {10.48550/ARXIV.2201.00049},
	urldate = {2022-08-23},
	author = {Somhorst, Frank H. B. and van der Meer, Reinier and Anguita, Malaquias Correa and Schadow, Riko and Snijders, Henk J. and de Goede, Michiel and Kassenberg, Ben and Venderbosch, Pim and Taballione, Caterina and Epping, Jörn. P. and Vlekkert, Hans. H. van den and Bulmer, Jacob F. F. and Lugani, Jasleen and Walmsley, Ian A. and Pinkse, Pepijn W. H. and Eisert, Jens and Walk, Nathan and Renema, Jelmer J.},
	year = {2022},
	note = {Publisher: arXiv
Version Number: 1},
}

@article{bib:perfect_quantum_error_correction_code,
  Title                    = {Perfect Quantum Error Correction Code},
  Author                   = {Raymond Laflamme and Cesar Miquel and Juan Pablo Paz and Wojciech Zurek},
  Journal                  = {Physical Review Letters},
  Year                     = {1996},
  Number                   = {198},
  Volume                   = {77}
}

@article{bib:coherent_error_surface_codes,
  Title                    = {Correcting coherent errors with surface codes},
  Author                   = {Sergey Bravyi and Matthias Englbrecht and Robert König and Nolan Peard},
  Journal                  = {npj Quantum Information},
  Year                     = {2018},
  Number                   = {1},
  Volume                   = {4},
  Pages                    = {55}
}

@article{madsen_quantum_2022,
	title = {Quantum computational advantage with a programmable photonic processor},
	volume = {606},
	issn = {0028-0836, 1476-4687},
	url = {https://www.nature.com/articles/s41586-022-04725-x},
	doi = {10.1038/s41586-022-04725-x},
	language = {en},
	number = {7912},
	urldate = {2022-07-10},
	journal = {Nature},
	author = {Madsen, Lars S. and Laudenbach, Fabian and Askarani, Mohsen Falamarzi. and Rortais, Fabien and Vincent, Trevor and Bulmer, Jacob F. F. and Miatto, Filippo M. and Neuhaus, Leonhard and Helt, Lukas G. and Collins, Matthew J. and Lita, Adriana E. and Gerrits, Thomas and Nam, Sae Woo and Vaidya, Varun D. and Menotti, Matteo and Dhand, Ish and Vernon, Zachary and Quesada, Nicolás and Lavoie, Jonathan},
	month = {6},
	year = {2022},
	pages = {75--81},
}

@article{richardson-extrapolation,
  title = {Efficient Variational Quantum Simulator Incorporating Active Error Minimization},
  author = {Li, Ying and Benjamin, Simon C.},
  journal = {Phys. Rev. X},
  volume = {7},
  issue = {2},
  pages = {021050},
  numpages = {14},
  year = {2017},
  month = {6},
  publisher = {American Physical Society},
  doi = {10.1103/PhysRevX.7.021050},
  url = {https://link.aps.org/doi/10.1103/PhysRevX.7.021050}
}

@article{quasi-probability,
  title = {Error Mitigation for Short-Depth Quantum Circuits},
  author = {Temme, Kristan and Bravyi, Sergey and Gambetta, Jay M.},
  journal = {Phys. Rev. Lett.},
  volume = {119},
  issue = {18},
  pages = {180509},
  numpages = {5},
  year = {2017},
  month = {11},
  publisher = {American Physical Society},
  doi = {10.1103/PhysRevLett.119.180509},
  url = {https://link.aps.org/doi/10.1103/PhysRevLett.119.180509}
}

@article{tann_quantum_2022,
	title = {Quantum {Remote} {Entanglement} for {Medium}-{Free} {Secure} {Communication}?},
	copyright = {Creative Commons Attribution Non Commercial No Derivatives 4.0 International},
	url = {https://arxiv.org/abs/2202.00830},
	doi = {10.48550/ARXIV.2202.00830},
	urldate = {2023-03-03},
	author = {Tann, Wesley Joon-Wie},
	year = {2022},
	note = {Publisher: arXiv Version Number: 1},
        journal={arXiv preprint arXiv:2202.00830},
}

@article{near-future,
  title = {Practical Quantum Error Mitigation for Near-Future Applications},
  author = {Endo, Suguru and Benjamin, Simon C. and Li, Ying},
  journal = {Phys. Rev. X},
  volume = {8},
  issue = {3},
  pages = {031027},
  numpages = {21},
  year = {2018},
  month = {7},
  publisher = {American Physical Society},
  doi = {10.1103/PhysRevX.8.031027},
  url = {https://link.aps.org/doi/10.1103/PhysRevX.8.031027}
}

@article{goodfellow2014generative,
  title={Generative adversarial nets},
  author={Goodfellow, Ian and Pouget-Abadie, Jean and Mirza, Mehdi and Xu, Bing and Warde-Farley, David and Ozair, Sherjil and Courville, Aaron and Bengio, Yoshua},
  journal={Advances in neural information processing systems},
  volume={27},
  year={2014}
}

@misc{tensorflow2015-whitepaper,
title={ {TensorFlow}: Large-Scale Machine Learning on Heterogeneous Systems},
url={https://www.tensorflow.org/},
note={Software available from tensorflow.org},
author={
    Mart\'{i}n~Abadi and
    Ashish~Agarwal and
    Paul~Barham and
    Eugene~Brevdo and
    Zhifeng~Chen and
    Craig~Citro and
    Greg~S.~Corrado and
    Andy~Davis and
    Jeffrey~Dean and
    Matthieu~Devin and
    Sanjay~Ghemawat and
    Ian~Goodfellow and
    Andrew~Harp and
    Geoffrey~Irving and
    Michael~Isard and
    Yangqing Jia and
    Rafal~Jozefowicz and
    Lukasz~Kaiser and
    Manjunath~Kudlur and
    Josh~Levenberg and
    Dandelion~Man\'{e} and
    Rajat~Monga and
    Sherry~Moore and
    Derek~Murray and
    Chris~Olah and
    Mike~Schuster and
    Jonathon~Shlens and
    Benoit~Steiner and
    Ilya~Sutskever and
    Kunal~Talwar and
    Paul~Tucker and
    Vincent~Vanhoucke and
    Vijay~Vasudevan and
    Fernanda~Vi\'{e}gas and
    Oriol~Vinyals and
    Pete~Warden and
    Martin~Wattenberg and
    Martin~Wicke and
    Yuan~Yu and
    Xiaoqiang~Zheng},
  year={2015},
  howpublished={\url{https://github.com/tensorflow/tensorflow}},
}

@misc{chollet2015keras,
  title={Keras},
  author={Chollet, Fran\c{c}ois and others},
  year={2015},
  publisher={GitHub},
  howpublished={\url{https://github.com/fchollet/keras}},
}

@article{mirza2014conditional,
  title={Conditional generative adversarial nets},
  author={Mirza, Mehdi and Osindero, Simon},
  journal={arXiv preprint arXiv:1411.1784},
  year={2014}
}

@article{pal2021torchgan,
  doi = {10.21105/joss.02606},
  url = {https://doi.org/10.21105/joss.02606},
  year = {2021},
  publisher = {The Open Journal},
  volume = {6},
  number = {66},
  pages = {2606},
  author = {Avik Pal and Aniket Das},
  title = {TorchGAN: A Flexible Framework for GAN Training and Evaluation},
  journal = {Journal of Open Source Software},
  howpublished={\url{https://github.com/torchgan/torchgan}},
}

@article{kingma2014adam,
  title={Adam: A method for stochastic optimization},
  author={Kingma, Diederik P and Ba, Jimmy},
  journal={arXiv preprint arXiv:1412.6980},
  year={2014}
}

\onecolumn
\appendix

\section*{Author Contributions}
Conceptualization, Krz.W. and Kam.W.; methodology, R.P. and Krz.W; validation, Kam.W.; formal analysis, R.P., Krz.W.; investigation, R.P. and Krz.W.; writing—original draft preparation, R.P. and Krz.W.; writing—review and editing, Krz.W.; visualization, R.P. and Krz.W.; supervision,
Kam.W and K.C.; project administration, K.C.; funding acquisition, K.C. and Kam.W.
All authors have read and agreed to the published version of the manuscript.

\section*{Funding}
The authors would like to acknowledge that this paper has been written based on the results achieved within the WrightBroS project. This project has received funding from the European Union’s Horizon 2020 research and innovation programme under the Marie Skłodowska-Curie grant agreement No 822483. Supplementarily, this research work has been co-financed from Polish financial resources for science in 2019-2023 conferred for implementation of the co-financed international project. 
Disclaimer. The paper reflects only the author's view and the Research Executive Agency (REA) is not responsible for any use that may be made of the information it contains.

\section*{Conflict of Interests}
The authors declare no conflict of interest.

\end{document}